\documentclass[%
 aps,
 showkeys,
 amsmath,amssymb,
 superscriptaddress, 
 floatfix,
]{revtex4-1}

\usepackage{graphicx}
\usepackage{dcolumn}
\usepackage{bm}
\usepackage{hyperref}

\let\cite\citep
\newcommand{\etal}[1]{} 

\usepackage[utf8]{inputenc} 
\usepackage{color}
\usepackage{etoolbox} 
\usepackage{gensymb} 
\usepackage{textgreek} 
\usepackage{bm} 
\newcommand{\um}{\textmu m} 
\newcommand{\uJ}{\textmu J} 
\newcommand{\Wcm}[2][0]{\ifnumcomp{0}{=}{#1}{}{$#1~\cdot$~}$10^{#2}$~W~cm$^{-2}$} 
\def\32omega{$\frac{3}{2}\omega$} 
\def\2omega{$2\omega$} 
\def\12omega{$\omega$/$2$} 

\begin{document}

\title{Relativistic electron acceleration by mJ-class kHz lasers normally incident on liquid targets}


\author{Scott Feister}
 \affiliation{Department of Astronomy and Astrophysics, University of Chicago, Chicago, IL 60615, USA}
 \affiliation{\mbox{Department of Physics, The Ohio State University, Columbus, OH, 43210, USA}} 
 \affiliation{Innovative Scientific Solutions, Inc., Dayton, OH, 45459, USA}

\author{Drake R. Austin}
 \affiliation{\mbox{Department of Physics, The Ohio State University, Columbus, OH, 43210, USA}} 
 \affiliation{Innovative Scientific Solutions, Inc., Dayton, OH, 45459, USA}
 
\author{John T. Morrison}
 \affiliation{Innovative Scientific Solutions, Inc., Dayton, OH, 45459, USA}

\author{Kyle D. Frische}
 \affiliation{Innovative Scientific Solutions, Inc., Dayton, OH, 45459, USA}

\author{Chris Orban}
 \affiliation{\mbox{Department of Physics, The Ohio State University, Columbus, OH, 43210, USA}} 
 \affiliation{Innovative Scientific Solutions, Inc., Dayton, OH, 45459, USA}
 
\author{Gregory Ngirmang}
 \affiliation{\mbox{Department of Physics, The Ohio State University, Columbus, OH, 43210, USA}} 
 \affiliation{Innovative Scientific Solutions, Inc., Dayton, OH, 45459, USA}
 
\author{Abraham Handler}
 \affiliation{\mbox{Department of Physics, The Ohio State University, Columbus, OH, 43210, USA}} 
 \affiliation{University of Dayton Research Institute, Dayton, OH, 45469, USA}

\author{Joseph R. H. Smith}
 \affiliation{\mbox{Department of Physics, The Ohio State University, Columbus, OH, 43210, USA}} 

\author{Mark Schillaci}
 \affiliation{\mbox{Department of Physics, The Ohio State University, Columbus, OH, 43210, USA}} 

\author{Jay A. LaVerne}
 \affiliation{Radiation Laboratory and Department of Physics, University of Notre Dame, Notre Dame, IN 46556, USA} 

\author{Enam A. Chowdhury}
 \email{chowdhury.24@osu.edu}
 \affiliation{\mbox{Department of Physics, The Ohio State University, Columbus, OH, 43210, USA}} 
 \affiliation{Intense Energy Solutions, LLC., Plain City, OH, 43064, USA}

\author{R.R. Freeman}
 \affiliation{\mbox{Department of Physics, The Ohio State University, Columbus, OH, 43210, USA}} 
 
\author{W.M. Roquemore}
 \affiliation{Air Force Research Laboratory, WPAFB, OH, 45433, USA}
 
\date{\today}

\keywords{Strong field laser physics; Electron radiation; Spectrometry; Experimental physics; Particle-in-cell}

\begin{abstract}
We report observation of kHz-pulsed-laser-accelerated electron energies up to 3~MeV in the $-\bm{k}_{\text{laser}}$ (backward) direction from a 3~mJ laser interacting at normal incidence with a solid density, flowing-liquid target. The electrons/MeV/s.r. $>$1~MeV recorded here using a mJ-class laser exceeds or equals that of prior super-ponderomotive electron studies employing lasers at lower repetition-rates and \emph{oblique} incidence. Focal intensity of the 40-fs-duration laser is $1.5 \cdot 10^{18}$~W~cm$^{-2}$, corresponding to only $\sim$80~keV electron ponderomotive energy. Varying laser intensity confirms electron energies in the laser-reflection direction well above what might be expected from ponderomotive scaling in normal-incidence laser-target geometry. This direct, normal-incidence energy spectrum measurement is made possible by modifying the final focusing off-axis-paraboloid (OAP) mirror with a central hole that allows electrons to pass, and restoring laser intensity through adaptive optics. A Lanex-based, optics-free high-acquisition rate ($>$100~Hz) magnetic electron-spectrometer was developed for this study to enable shot-to-shot statistical analysis and real-time feedback, which was leveraged in finding optimal pre-plasma conditions. 3D Particle-in-cell simulations of the interaction show qualitative super-ponderomotive spectral agreement with experiment. The demonstration of a high-repetition-rate, high-flux source containing $>$MeV electrons from a few-mJ, 40~fs laser and a simple liquid target encourages development of future $\geq$kHz-repetition, fs-duration electron-beam applications.
\\
\end{abstract}





\maketitle









\section{Introduction}
High intensity lasers ($>$\Wcm{17}) have enabled particle acceleration to high energies (keV to GeV) through a variety of methods and mechanisms, all of which exploit the strong electromagnetic fields that exist during intense laser plasma interactions. Some high intensity laser sources are capable of accelerating electrons to GeV energies through direct laser-field and/or laser wakefield acceleration \cite{esaray2009}. In addition, ultra-intense laser-plasma interactions can create bright X-ray sources from high harmonic generation \cite{Dromey2006}, bremsstrahlung collisions of electrons \cite{murnane1991ultrafast}, betatron oscillations \cite{kneip2010}, Compton scattering \cite{phuoc2012, chen2013} and other techniques. These applications benefit from advances in the peak power that ultra-intense laser systems can provide.

While these experiments are very interesting from a scientific perspective, for laser technology to provide a compelling alternative to conventional accelerators, and for accelerator-based light sources, it is an important goal for the field to produce ultra-intense laser pulses with high \emph{average} power \cite{hooker2013}. This involves creating intense laser pulses at a high repetition rate. Most laser facilities operate with a repetition rate between 0.03-1 Hz \cite{hooker2013}, which is orders of magnitude less than conventional accelerators. Moreover, ultra-intense laser experiments typically operate in a single-shot mode with diagnostics designed for single-shot operation. Next generation laser facilities are soon coming online, with laser systems boasting 10 Hz repetition rate at peak powers exceeding 1 PW \cite{rus2015}. However, these systems would still need high repetition rate target insertion and alignment systems where inexpensive targets can be brought in and aligned within a few ms between laser shots, and experimental diagnostics to extract data from such interactions at the operating duty cycle.

Millijoule-class tabletop short-pulse lasers, with reduced size, complexity, cost, and operating expenses, can provide a useful template for the next generation laser experimental systems to follow. Such a laser experimental system is discussed in this paper, which can operate at a high repetition rate (up to $\sim$1~kHz).  One example of 0.5~kHz laser-plasma electron acceleration is found in \etal{He}\citet{he2013}, with acceleration to 100~keV by laser-wakefield acceleration and demonstrated applications in ultrafast electron diffraction \cite{he2016}. Another high-repetition-rate laser wakefield acceleration study by \etal{Guénot}\citet{guenot2017} used a near-single-cycle laser pulse with temporal contrast better than 10$^{10}$ focused into an underdense flowing gas jet target to accelerate electrons to $\sim$5~MeV. Experiments with these mJ-class laser systems can produce GeV/cm accelerating fields and energetic bursts of electrons with MeV energies \cite{milchberg2015}.
This paper will present direct evidence for some of the most energetic, highest-efficiency MeV electron acceleration that has been observed in the literature to date from laser systems of this kind.

Laser-plasma interaction experiments at and below relativistic threshold intensity (\Wcm{18} for 800~nm light) typically show exponential decays in electron spectra with characteristic energies well below 100~keV (e.g. \cite{mordovanakis2010}), but some studies using dense targets with a controlled pre-plasma have shown higher-energy electron acceleration in oblique-incidence laser-target configurations. \etal{Mordovanakis}\citet{mordovanakis2009} used a 3~mJ laser to accelerate 7~pC of charge to between 100~keV and 1.4~MeV with an energy spectrum peaked at 0.8~MeV, and \etal{Mao}\citet{mao2012} used a 240~mJ laser to accelerate, in at least a 14\degree{} full-angle cone, 100~pC of electrons per shot with similar energy. \etal{Bocoum}\citet{bocoum2016} explored the role of pre-plasma scale length in electron confinement within the target, accelerating electrons with up to 600~keV energy in an estimated 11~pC bunch after careful optimization of pre-plasma scale length ($L\sim0.1\lambda$). In prior studies, the ejection of energetic electrons depends on \textit{p}-polarized laser pulses and large angles of incidence ($\geq$45\degree{}). 
 Electrons tend to be emitted near the laser specular and normal directions with laser-to-electron energy conversion efficiency below 0.05\%. The conversion efficiencies in laser-wakefield and dense gas jet experiments likewise tend to fall below this level. A recent study by \etal{Salehi}\citet{salehi2017} demonstrated significant numbers of forward-going $>1$~MeV electrons from 4~mJ to 10~mJ laser interactions with a dense gas jet target. According to their measurements, the laser-to-electron conversion efficiency in those experiments was roughly 0.015\% or less.
 
In the presently studied case of normal-incidence laser-target interaction, \textit{p}-polarization-dependent electron acceleration mechanisms are suppressed. However, three recently published experimental and computational papers \cite{morrison2015backward, orban2015backward,ngirmang2016} have demonstrated how a wide spray ($\pm$40\degree{} cone) of electrons can be accelerated to $>$120~keV with $\gtrsim$1.5\% laser-to-electron energy conversion efficiency in the laser reflection direction for few-mJ, \Wcm{18}, normal-incidence interactions. A standing wave and direct laser acceleration theoretical framework was developed in \etal{Orban}\citet{orban2015backward} to provide a physical interpretation of this normal-incidence, extremely-high-efficiency electron acceleration by a mJ-class relativistic laser. Simulations in that paper showed $\gtrsim$0.5\% efficiency for electrons accelerated to $>$MeV through the same mechanism, without experimental confirmation.

This paper contributes two scientific advancements to the study of laser-based electron acceleration. First, $>$MeV electrons are experimentally observed in the explicit $-\bm{k}_{\text{laser}}$ (backward) direction for a normal-incidence laser-plasma-interaction, despite a $81$~keV relativistic ponderomotive energy \cite{wilks1992}. Until the present work there has been no direct experimental confirmation of electrons with $>$120~keV energies in experiments of this geometry and directionality. Second, the measured electrons/MeV/s.r. above 1~MeV generated using this low-energy laser (3~mJ) exceeds or equals that of related prior studies employing lasers at oblique incidence of similar intensity. This paper provides the first clear evidence, through direct electron detection in a magnetic spectrometer, that the highly efficient interactions of \etal{Morrison}\citet{morrison2015backward} produce large numbers of $>$MeV electrons. 

This work also contributes novel methods for high-acquisition-rate High Energy Density (HED) experiment and normal-incidence laser-target setup. Due to the possibility of damaging delicate optical systems, it is difficult to find studies with ultra-intense laser interactions at normal incidence in the literature, and, perhaps due to the difficulty of obtaining a direct line-of-sight between an electron spectrometer and the mirror opening, it is even harder to find studies with careful investigation of electron acceleration in this configuration.  
According to the authors' best knowledge, a 0\degree{} (direct laser back-reflection direction) measurement of super-ponderomotive electron energy in a normal-incidence laser configuration has no precedent in the literature. In this paper, an experimental electron energy characterization is presented which confirms electron acceleration up to 3~MeV in the laser back-reflection direction.   These measurements are performed while taking advantage of the kHz repetition rate of a mJ-class ultra-intense laser system. In the field of HED ultra-intense laser-matter interactions, shot-to-shot variations can frustrate experiments on low-repetition-rate laser systems where, by necessity, researchers are often forced to draw conclusions from only a few shots. A mean spectrum and standard deviation range using fifty electron spectra (each an average of ten shots) is presented, which was adequate for capturing the range of shot-to-shot variations in this experiment.
This energy characterization is discussed with new 3D Particle-in-cell (PIC) simulations and in a theoretical framework of normal-incidence, laser reflection super-ponderomotive electron acceleration at relativistic threshold intensity.

\section{Experimental Setup}

\begin{figure}[htbp]
    \centering
    \includegraphics[width=8cm]{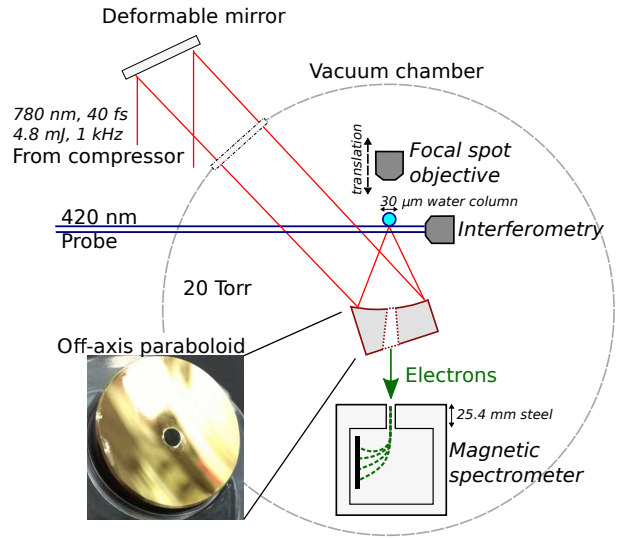}
    \caption{\label{fig:setup}To measure the electron spectrum in the direction anti-parallel to the laser axis (back-reflection direction), a 3~mm hole was drilled into the 25.4~mm diameter 30\degree{} gold-coated f/1.1 OAP. The central portion of specular-direction electrons, accelerated by normal-incidence laser interactions with a flowing water target, pass through the hole. Electron energies are characterized by a magnetic spectrometer located 127~mm behind the target. The laser is polarized in the plane of the diagram.}
\end{figure}

The target area and experimental setup for this experiment are sketched in Fig. \ref{fig:setup}. A 1~kHz Ti:sapphire laser system delivers 4.8~mJ (2.9~mJ on target) with a pulse duration of 40~fs full-width at half-maximum (FWHM). The laser is focused onto a high-velocity water stream (a column 30~\um{} in diameter) \cite{morrison2015backward}, which enables the study of laser-plasma interactions at 1~kHz. The laser is focused using an off-axis paraboloid (OAP). In order to make the desired normal-incidence $-\bm{k}_{\text{laser}}$ (backward) experimental measurement, a 3~mm hole was drilled into its center. As shown in Fig.~\ref{fig:setup}, this hole allowed a portion of electrons accelerated from the target in a 6.3\degree{} full-angle cone in the laser reflection direction to pass through to an electron spectrometer placed 127~mm behind the target. 

\begin{figure}[htbp]
    \centering
    \includegraphics[width=8cm]{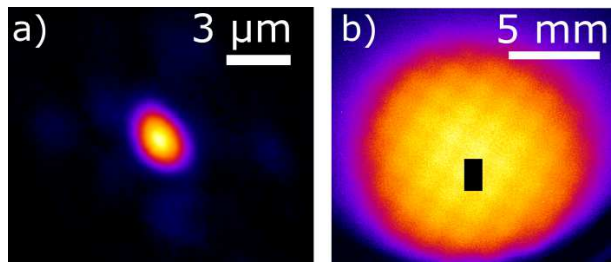}
    \caption{\label{fig:spots}
    a) The focal spot measured after deformable mirror optimization, with a calculated \Wcm[1.5]{18} intensity. b) Portion of the electron beam that passes through the hole of the OAP, imaged with a Lanex phosphor camera \cite{orban2015backward}. The black rectangle denotes the region (1~mm wide x 1.7~mm tall) that enters the magnetic spectrometer's slit and is energy-resolved. In this image, a diagnostic 250~\um{} thick stainless steel mesh sits in front of the camera to check spectral uniformity; both camera and mesh are removed before electron spectra measurements.
    }
    \end{figure}

Modifying the OAP in this way to obtain such a measurement is atypical in experimental high-intensity laser-plasma interaction physics because it severely distorts the laser focus, reducing on-target intensity. An adaptive optics system, similar to the one pioneered by \etal{Albert}\citet{albert2000}, was custom designed and configured to compensate for intensity degradation due to the hole in the OAP. The deformable mirror was inserted into the laser chain after the laser compressor. The in-vacuum focal spot area after the OAP was measured using a microscope objective imaging system and CCD camera. The imaged laser focal spot area and quality was analyzed in real-time to close a feedback loop (at 60~Hz) with deformable mirror actuators through a genetic algorithm (written by the authors in LabVIEW).
Despite the energy loss and decreased optical quality due to the hole in the OAP, the adaptive optics setup converged within five minutes to a 2.2~\um{} FWHM in-vacuum focal spot, with 2~mJ of the 2.9~mJ on-target energy contained in the Gaussian laser focus (see Fig. \ref{fig:spots}(a)). The pulse energy stability was $3\%$ peak to peak with $<1\%$ rms.
Accounting for the hole and adaptive optics corrections, the total on-target laser energy in this experiment was 2.9~mJ and laser intensity was \Wcm[1.5]{18}.

\begin{figure}[htbp]
    \centering
    \includegraphics[width=13cm]{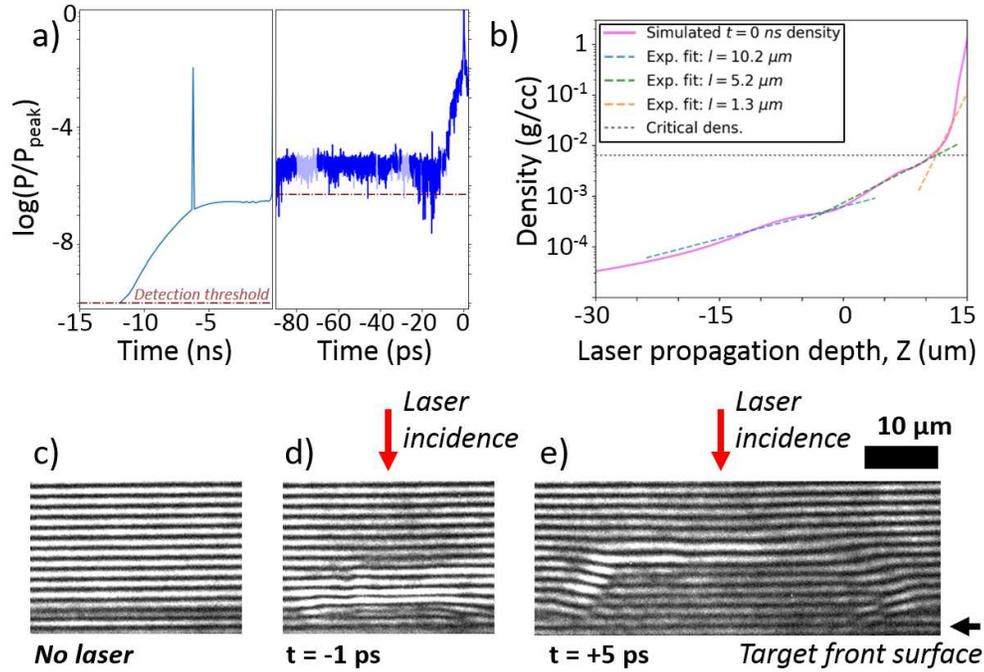}
    \caption{\label{fig:ppulse}a) Measured characteristic pre-pulse intensity on the nanosecond and picosecond timescales. The spikes in intensity at $t$~=~-6.2~ns and $t$~=~-15.3~ps are 40-fs-duration main-pulse replicas. The main pulse arrives at t~=~0~ps. Light blue segments of the picosecond trace have been modified to remove third-order cross-correlation artifacts. b) Pre-plasma density profile from the long-pulse component only of (a) as modeled in FLASH. Ablated material extends tens of microns from the target, with a steeper profile nearer to solid density (1~g/cc). Original target interface is at Z=0~\um{}. The density profile can be described by multiple exponential decays, indicated with their characteristic scale lengths; long scale lengths exist far from target. The 800~nm plasma critical density is indicated assuming eight free electrons per H$_2$O molecule (a typical value derived from the PIC simulations, due to ionization by the main pulse). c) Reference interferometric image for (d) and (e), with the laser blocked. d) Femtosecond-gated 420~nm probe beam interferometry captures some information about the target condition 1~ps before main pulse ionization. Fringe shifts are evident up to several microns away from the target, demonstrating the presence of pre-ablated material. According to the fringe shift direction, ionization is incomplete at this stage, which may affect the sensitivity of the interferometry to lower-density material further from the target. e) Interferometry delayed to 5~ps after main pulse ionization. The direction of the fringe shift indicates that the material is in the ionized state. Fringe shifts at tens of microns away from the target indicate the presence of material at this distance. Due to the brief time interval between the two interferometric measurements this is interpreted as diffuse material that was present before the ultra-intense laser interaction, but undetectable in interferometry due to its low density and partially ionized state.}
\end{figure}

The laser's intrinsic pre-pulse, which extends over several nanoseconds, is experimentally manipulated using electro-optic gating.  Pockels cell timing and extinction ratio are adjusted at four points throughout the laser amplification chain to reach the condition of high-number/high-energy electron acceleration presented in this paper. These adjustments are made in response to electron spectra that are acquired, analyzed and displayed in real-time. As a consequence of $\sim$3~ns rise time and temporal jitter of the Pockels cells, combined with the unsaturated gain of the pre-pulse in the amplifiers, the photodiode-measured pre-pulse energy stability was 15\% rms. Importantly, the optimal conditions do \emph{not} correspond to the highest level of pre-pulse suppression that the system is capable of. 
Nanosecond-timescale laser contrast was measured using a Thorlabs DET10A photodiode and a 2.5~GHz oscilloscope. Dynamic range in the measurement was enhanced by recording and recombining multiple oscilloscope measurements, each with different levels of neutral-density optical filtering on the photodiode \cite{morrison2015backward}.  Due to the oscillator pulse-train and ghost reflections within the multi-pass laser amplifiers,  40-fs-duration short-pulse replicas are expected at fixed delay times in the nanoseconds preceding the main pulse. Evidence for these short-pulse replicas can be inferred from the nanosecond signal, and deconvolved using the impulse response of the photodiode/oscilloscope system. A scanning third-order cross-correlator measurement of the picosecond-timescale laser contrast confirms this conclusion with greater temporal precision.
The laser amplified spontaneous emission (ASE) contrast for this experiment is measured to be $5~\cdot~10^{-6}$ at {$t=-15$~ps} with a 40-fs-duration short pulse replica at $t=-6.2$~ns containing a shot-to-shot averaged 70~\uJ{} at \Wcm{16}. A plot of the average laser pre-pulse profile is shown in Fig. \ref{fig:ppulse}(a). As mentioned earlier, these experiments could have been performed with more aggressive suppression of the $t=-6.2$~ns short pulse replica and the ASE. Instead, a relatively low-contrast condition was adopted because this empirically maximizes the number of high-energy electrons accelerated in the laser reflection direction.

To characterize the effect of the pre-pulse on the condition of the target, a hydrodynamic simulation of pre-plasma evolution was performed using the FLASH code \cite{flash2000, flash2009, flash2012}. The simulation was performed in 2D R-Z cylindrical geometry. 
Because of the 2D cylindrical geometry, the water column target is represented as a water droplet of radius 15~\um{}. The simulation extent is $R=\pm$28~\um{} and $Z=$-85~\um{} to $Z=+$35~\um{}, with $R=Z=0$ at the front center surface of the water droplet. Cell spatial resolution is 0.25~\um{}. The timestep varies during the simulation according to the Courant limit, which is determined by the fastest sound speed in the simulation and the cell size. The laser focus convergence with the experimental f/1.1 and focal depth is also set to match the experiment (2.2~um{} Gaussian focus located 15~\um{} beneath the target surface). Because highly intense femtosecond-timescale laser-plasma-interaction physics can not be accurately modeled in a hydrodynamic simulation, the 40~fs replica pulses at $t$=-6.2~ns and $t$=-15.3~ps are excluded from the pre-pulse profile. Results of this simulation show that even without these features, a pre-plasma with tens-of-microns extent develops (Fig.~\ref{fig:ppulse}(b)).

Laterally sheared optical interferometry is used in Fig. \ref{fig:ppulse}(d) and \ref{fig:ppulse}e to show target state near the time of interaction. To maintain order-femtoseconds synchronicity between pump and the probe pulses, seed pulses from the same ultrafast oscillator are used for generating pump and probe. Separate amplifier systems are used. The amplified probe beam is frequency-shifted and frequency doubled, resulting in 420~nm probe which is optically isolated from the on-target main pulse second harmonic light (390~nm) using a narrow-linewidth optical filter (see \etal{Feister}\citet{feister2014}). The probe beam passes through the interaction region, as shown in Fig. \ref{fig:setup}, then enters a microscope objective for imaging purposes. The beam path is then is split for separate shadowgraphic and interferometric analysis. The interferometric beam path undergoes lateral shearing such that the vacuum region of the beam image is overlapped with the interaction region to create interference fringes, which change in angle based on integrated index of refraction differences between the overlapped regions.

A comprehensive characterization of electron energies in the laser-plasma-interaction was desired, and stochastic shot-to-shot fluctuations are a typical feature of ultra-high intensity laser plasma interactions with near solid density targets. To capture stochastic shot-to-shot fluctuations in accelerated electron energy, a high acquisition rate electron spectrometer was designed, built, and calibrated for this experiment. The spectrometer consists of a 1~mm slit followed by two yoked magnets to deflect electrons according to their energy, and encased in a steel enclosure. The electrons are imaged using a Tb-doped $\mathrm{Gd_2 O_2 S}$ phosphor screen (Lanex regular) that is placed directly upon a bare linear CCD chip (MighTex CCD camera, Toshiba 1304DG linear CCD) and made light-tight with a 25~\um{} thick covering of aluminum. An absolute calibration of this detector, in electrons per CCD count (adjusted for CCD bias) at 1~MeV, 2.0~MeV, and 2.5~MeV, was obtained using the University of Notre Dame Radiation Laboratory Van de Graaff 3~MeV electron accelerator.
Considerations were taken of the continuous-source nature of the Van de Graaff electron beam and the discretely pulsed nature of the experimental electron beam. The Lanex regular phosphor response is linear with electron number at these fluences, with a 1/e light decay time of 0.6 ms \cite{Nikl2006}. For a continuous source, the total light emitted from the screen during a single CCD exposure of fixed, arbitrarily short duration will include some decaying phosphorescent emission of electrons incident prior to CCD exposure, but this light will integrate equally to the missed phosphorescent decay of electrons incident during CCD exposure. Although there are shot-to-shot variations in the laser experiment, the same argument holds true for the 10~ms experimental exposures that were used to gather data for Figs.~\ref{fig:espec}-\ref{fig:specscan}. Due to the kHz repetition rate these 10~ms exposures capture integrated light from ten consecutive electron pulses.
Because the exposure time is much longer than the decay time, a shot occurring prior to exposure is unlikely to affect the recorded exposure by more than $\sim$2\%, even with 100\% deviation from the mean. Moreover, this effect should neither bias the measured fluence up or down, as mentioned earlier. 
An MCNP model of the calibration setup at the Van de Graaff accelerator was used to correct for electron scattering within the beam's aluminum exit window from vacuum, the at-air environment of the spectrometer, and the Van-de-Graaff-facility Faraday cup. The absolute experimental result at these discrete energies, after corrections from MCNP modeling of the effects just mentioned, were used to determine the electron spectrometer response characteristics applied to generate the experimental electron spectra of Figs.~\ref{fig:espec} and \ref{fig:specscan}. The electron spectrometer's electron energy positions (at the detector) were verified in two ways. First, Hall probe magnetic field measurements were combined with relativistic particle deflection calculations. Second, energy positions were verified between 1~MeV and 2.5~MeV using the Van de Graaff accelerator.

Before performing energy characterization of back-reflection direction electrons, the laser focus was aligned to the target as follows: the cylindrical water stream was vertically translated such that the laser was incident 1~mm below the nozzle output, vertically angled for normal incidence (as seen in shadowgraphy/interferometry, through air breakdown by the laser above 150~Torr), horizontally translated for normal incidence (as seen by optical back-reflection), and focally translated for strongest backward-going electron signal on the electron spectrometer. The in-vacuum laser pointing and target position stability was measured to be $\sim$1~\um{}. The electron spectrometer is located 127~mm behind the target (100~mm behind the OAP). Before measurements, a separate permanent magnet is swung into place between the OAP and electron slit (magnetically but not physically blocking the electrons' paths) to check that recorded signal drops to zero. This ensures the measured signal is from electrons and not due to stray X-rays or laser light. The spectrometer is triggered and acquires electron traces at $\sim$100~Hz.

Figure~\ref{fig:espec} shows experimental and 3D Particle-in-cell (PIC) simulation results for laser-reflection direction electron acceleration. These 3D PIC simulations of femtosecond timescale laser-matter interactions were performed using the Large-Scale Plasma (LSP) code  \cite{Welch2004}. The incident laser parameters were similar to that of the highest energy experimental results shown: intensity \Wcm[1.5]{18}, laser wavelength $\lambda = 800$~nm, Gaussian spot diameter 2.3~\um{}, and 30~fs FWHM pulse duration. The simulation space was 35~\um{} along the longitudinal ($k$-vector) direction, 40~\um{}~$\times~$40~\um{} in the transverse directions, and the cell size for these simulations was $\lambda /8 = 100$~nm with 27 macro-particles per cell per species (electrons, Oxygen, and Hydrogen ions). The density profile was assumed to be exponential with a 1.5~\um{} scale length. Due to computational limitations, the long-scale-length low-density component of the FLASH simulations was excluded. The simulations were run for more than 100~fs after the peak of the pulse reached the critical density surface of the pre-plasma. Backward-going electron spectra were obtained by tracking electron macroparticles and binning the charges and energies of these particles exiting the simulation space.

\section{Results and Discussion}
\begin{figure*}[htbp]
    \centering
    \includegraphics[width=13cm]{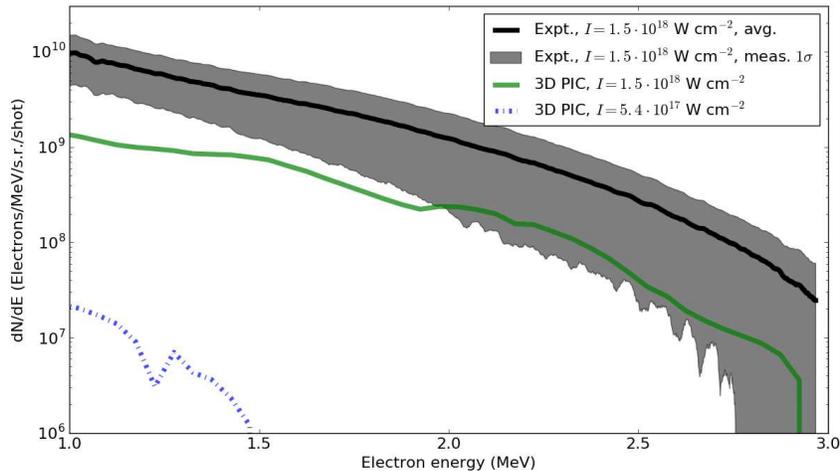}
    \caption{\label{fig:espec}
    Energy characterization of electrons accelerated in the laser reflection direction for normal-incidence laser interaction. The y-axis (charge density) for the experimental data is absolutely calibrated and can be directly compared to the 3D PIC simulation results shown. This calibration also includes an adjustment for detector sensitivity, number of shots per spectrum, and electron scattering losses in the 20~Torr chamber pressure. The black curve shows the mean of fifty consecutive experimental electron spectra in which the in-vacuum laser intensity was measured to be \Wcm[1.5]{18}. The standard deviation range of the measured spectra brackets the curve in gray. 3D PIC electron spectra are plotted for two intensities, \Wcm[1.5]{18} and \Wcm[5.4]{17}, which highlights the rapid turn-on with intensity of this acceleration mechanism. For both PIC simulations, the pre-plasma density profile is the same and only electrons emitted in the laser reflection direction within a 6.3\degree{} full-angle cone are included.
    }
    \end{figure*} 

\begin{figure*}[htbp]
    \centering
    \includegraphics[width=13cm]{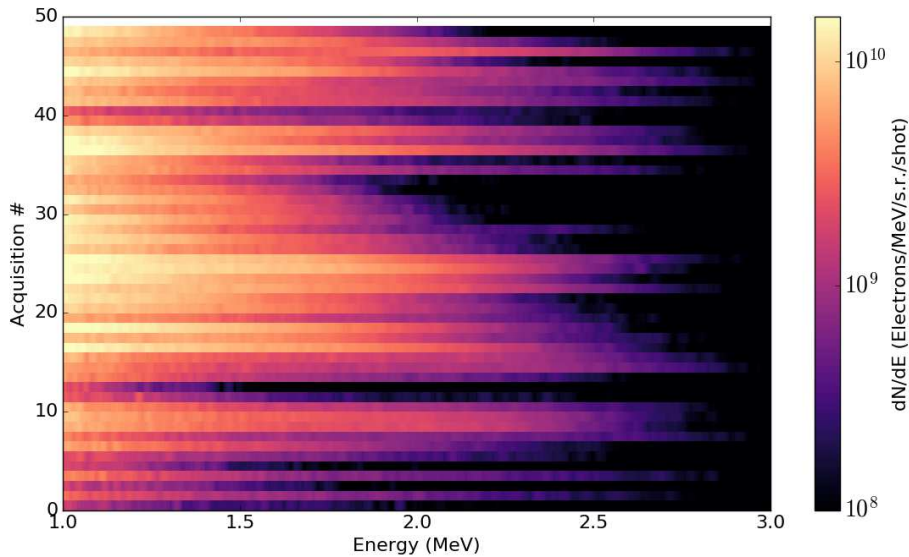}
    \caption{\label{fig:multishot} A waterfall plot presents fifty consecutive experimental electron spectra from \Wcm[1.5]{18} laser interactions used to compute the mean electron spectrum of Fig. \ref{fig:espec}. These measurements are enabled by the union of a high-repetition-rate laser, liquid targets, electronic detectors, and an automated acquisition system. Each spectrum, represented by an individual stream in the plot, is the result of a ten-millisecond CCD exposure (averaging over ten laser-target shots). This plot in total represents 0.5 seconds of continuous data acquisition at a 100~Hz rate. Large variations in the electron spectra are partly due to fluctuations in the laser pre-pulse and the sensitivity of the ultra-intense laser-matter interactions with near solid density targets to the properties of the pre-plasma.
    }
    \end{figure*}

\begin{figure*}[htbp]
    \centering
    \includegraphics[width=13cm]{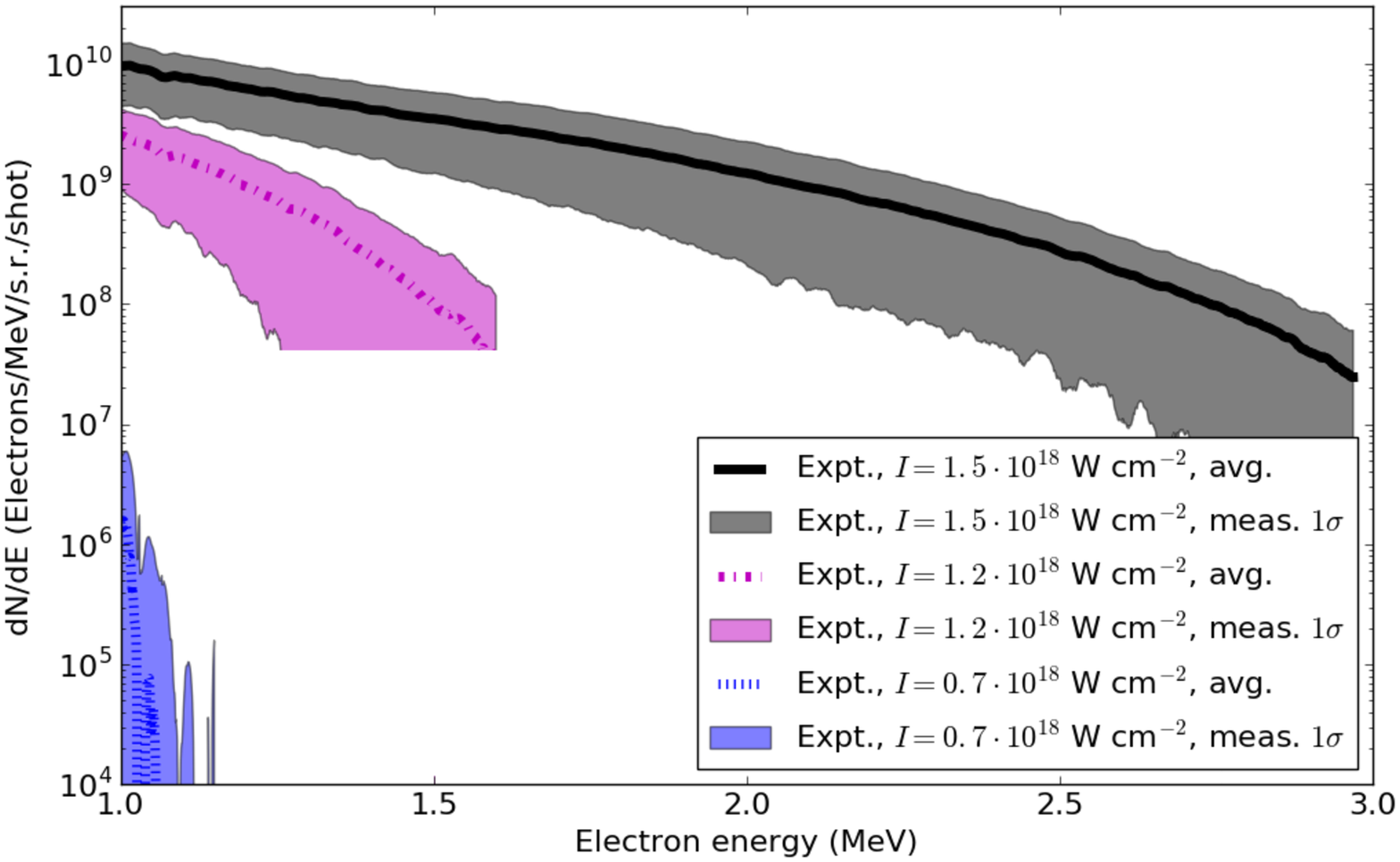}
    \caption{\label{fig:specscan} Experimental electron energy characterization for three laser intensities shows extremely steep intensity scaling. Intensity is reduced through use of a variable waveplate and polarizer, meaning the main pulse and its pre-pulse are equally attenuated. Fifty to one hundred consecutive electron spectra are acquired at each laser intensity, and a mean electron spectrum (solid line) as well as the standard deviation range (1$\sigma$ range, filled area) are plotted. As in Fig. \ref{fig:espec}, the experimental spectra are adjusted for detector sensitivity, number of shots per spectrum, and electron scattering losses due to the 20~Torr chamber pressure.}
    \end{figure*}
    
The data shown in Fig.~\ref{fig:espec} represents the synthesis of high repetition rate targetry, a high repetition rate intense laser and high acquisition rate (100~Hz) spectrometery. There are several features of Fig.~\ref{fig:espec} that merit discussion. First, due to the low repetition rate of most ultra-intense laser systems \cite{hooker2013},
it is relatively rare in studies of ultra-intense lasers interacting with dense targets to see plots with both a mean and standard deviation for electron spectra.  Each shaded infinitesimal area slice between a particular electron energy E\textsubscript{0}, and E\textsubscript{0} + $\Delta$E in Fig.~\ref{fig:espec} is determined from 50 spectral acquisitions (500~laser-interactions with the target). This capability is an important advantage of kHz ultra-intense laser systems because, as shown in Fig.~\ref{fig:multishot}, one finds large variations in the electron spectrum in laser plasma interactions with near solid targets. A major contribution to electron spectrum fluctuations is the sensitivity of ultra-intense laser interactions with solid-density targets to the properties of pre-pulse-generated pre-plasma \cite{bocoum2016}.  Few-shot experiments with low repetition rate ultra-intense laser systems should be similarly sensitive to these details. Secondly, the two higher show electrons accelerated to well above the ponderomotive scaling energy (81~keV for \Wcm[1.5]{18}) in large numbers. Results like these, which are insightful for understanding the limits of plasma acceleration processes,  may go unnoticed by few-shot experiments.

Third, compared with electron acceleration studies at oblique laser incidence, the measurements of  electrons/MeV/s.r. above 1.3~MeV for \Wcm[1.5]{18} exceed all previous observations in the literature for a few-mJ laser system. While there is tension between this measurement and the electron spectrum for the 3D PIC simulation, with the experiment indicating more electrons than expected from simulations, both the simulations and experiment qualitatively agree that large numbers of MeV electrons are ejected from the target. Figure~\ref{fig:espec} provides the first direct evidence that significant numbers of electrons can be accelerated to super-ponderomotive energies in water jet experiments with $\sim$\Wcm{18} intensities at near normal incidence. Although the experimental conditions in \etal{Morrison}\citet{morrison2015backward} do differ in some ways from the experimental conditions examined here, Fig.~\ref{fig:espec} also confirms an indirect measurement of the ejected electron energies from \etal{Morrison}\citet{morrison2015backward}.

Figure~\ref{fig:specscan} demonstrates the intensity sensitivity of the electron acceleration in these experiments. Earlier, it was shown in Fig.~\ref{fig:espec} that a 3D PIC simulation at \Wcm[5.4]{17} resulted in orders of magnitude lower numbers of ejected electrons despite laser intensity lowered by only about a factor of two from that of the previous one at \Wcm[1.5]{18}. In Fig.~\ref{fig:specscan} it is likewise found that, experimentally, the efficiency of the electron acceleration process is highly sensitive to the peak intensity in this range. 

An important aspect of the experiments presented here is that to avoid freezing the water jet target, the target chamber pressure could only be reduced to 20~Torr. A natural question is whether this affected the properties of the ultra-intense laser pulse prior to reaching the target. An empirical test was performed without a water jet target to examine this possibility. The laser was allowed to propagate through background pressures between 1~Torr and 100~Torr, and the spectrum and beam mode after going through focus is observed. No spectral shift was apparent below 40~Torr, and no beam mode degradation was observed below 80~Torr, indicating that self-focusing was insignificant at the 20~Torr operating pressure of the water jet experiments presented here. 
\textit{Geant4} \cite{agostinelli2003} modeling was also performed to assess whether electrons (especially at lower energies) would be deflected out of the electron spectrometer's acceptance angle due to scattering in ambient water vapor in the vacuum chamber.
The \textit{Geant4} simulations indicate that the 20~Torr ambient pressure in the target chamber diverts $\sim$22\% of $\sim$1~MeV electrons and 9\% of 2~MeV electrons from entering the spectrometer. These corrections were applied to the electron spectra plotted in Figs. \ref{fig:espec} and \ref{fig:specscan}. 

A theoretical basis for the super-ponderomotive energies of backward-going electron acceleration for this experiment is described in \etal{Orban}\citet{orban2015backward}. A standing wave is established between the forward-going laser and its backward reflection. As a result, the peak electric and magnetic fields significantly exceed the values present in the incident laser pulse due to constructive interference.
Unlike oblique incidence standing waves, as described for example by \etal{Mordovanakis}\citet{mordovanakis2009}, normal incidence standing waves have stable node and anti-node positions, which enhances electron acceleration by developing quasi-static charge separations in a process known as ponderomotive steepening \cite{estabrook1983}. The standing wave provides a pathway for electron acceleration both into and away from the target \cite{kemp2009,orban2015backward}, a process that can be influenced by quasi-static fields that arise from the charge separation caused by the standing wave. Importantly, electrons accelerated into the target will continue without another strong interaction with the laser field, while those accelerated away from the target (laser back-reflection direction) can be injected into the reflected light and undergo direct laser acceleration to higher energies. This direct laser acceleration of ejected electrons by reflected laser light provides a compelling explanation for the MeV electron energies detected experimentally in the current study. The appendix of \etal{Orban}\citet{orban2015backward} also provides an argument for why this mechanism should be much less efficient for laser intensities significantly below \Wcm{18}.

As has been discussed, the measured values for the number of electrons/MeV/s.r. are large compared to prior literature using similar laser energies and intensities. The measured values are also large compared with the 3D PIC simulation presented earlier, and it is argued here that this is due to the limitations of the pre-plasma extent adopted in the simulation. In particular, a multi-exponential-scale-length pre-plasma was not included in the LSP simulation. Instead, a single 1.5~\um{} scale length was assumed which is similar to the $\sim$1.3~\um{} exponential scale length that the FLASH simulation results in Fig.~\ref{fig:ppulse}(b) indicate is present near the critical density. Since many prior studies have emphasized the importance of the characteristic exponential scale length of the pre-plasma near the critical density this is an important aspect of the laser-matter interactions to include.

Although the FLASH results also indicate a $\gtrsim 5$~\um{} scale length extending from the target and indications from interferometry also point to the existence of an extended pre-plasma, 3D PIC simulations with a more extended pre-plasma were not presently possible due to computational expense. This affected the results of the PIC simulations in important ways. It is known from performing 2D PIC simulations using pre-plasmas that are similar to the FLASH density profile in Fig.~\ref{fig:ppulse}(b) and similar laser pulse parameters that the presence or absence of a low density, long-scale length component of the pre-plasma can have a strong effect on the number of MeV back-directed electrons. In a test by the authors, the addition of a diffuse (sub-sixth-critical-density), long-scale length pre-plasma with a scale length of 7.2~\um{} increased the conversion efficiency of back-directed $>120$~keV electrons from 0.67\% to as much as 3.4\%, an increase by a factor of five\cite{feisterDPP2016}. This dramatic increase in efficiency can be explained by the "capacitor" model developed by \etal{Link}\citet{link2011}, which considers in detail the process of electron ejection from short-pulse laser illuminated targets that are charging up as energetic electrons depart. As these electrons leave the electric potential on the surface of the conducting plasma target increases, which strongly affects the energy threshold for escape and the final energy of escaped electrons. \etal{Link}\citet{link2011} also considers the effect of target size, concluding "More electrons escape for internal electron distributions with more energy, for electron distributions with higher hot electron temperatures, and for larger targets." The simple reason for this being that larger conductors naturally have a lower electric potential, and due to the generically steep decrease in the numbers of energetic electrons with increasing energy in the target, a lower electric potential will allow many more electrons to escape. \etal{Link}\citet{link2011} discusses this in the context of multi-Joule laser systems, but the essential plasma physics is the same in describing mJ-class experiments. It should be emphasized that the size of the target depends on the spatial extent of the pre-plasma, and that 2D PIC simulations confirm the essential reasoning that increasing the target size in this way can lead to increases in the number of MeV scale electrons by a large factor. Our 3D PIC simulations therefore describe the ejection of electrons via a standing-wave-acceleration process, but the details of electrostatic turnaround and charging of the target are mis-represented in a way that reduces the energy and, importantly, number of MeV scale electrons in important ways. Consequently, the 3D PIC simulations confirm the existence of significant numbers of MeV scale electrons but the simulations underpredict the measured numbers of ejected electrons by a large factor.

It is interesting to consider the properties of the 6.3\degree{} full-angle divergence sub-beam (shown in Fig.~\ref{fig:spots}(b)) which is created by clipping through the OAP hole.
These estimates do not include the electrons in the sub-beam with energies below 1~MeV.
Integrating the fifty electron spectral curves (for \Wcm[1.5]{18}) from 1~MeV to 3~MeV and over the solid angle 9.6~msr of the OAP hole, one finds in the 500-shot-mean an acceleration of 6.9~pC/shot and 0.33\%~efficiency into $>$MeV electrons in the sub-beam. Since the laser-plasma interaction occurs continuously at 1~kHz, this represents a time-averaged 6.9~nA current of $\gtrsim$MeV electron bursts powered by a 3~W average-power laser. The liquid target recovers in tens of microseconds \cite{feister2014}, so a higher-repetition-rate laser could be expected to generate linearly higher currents. The highest-efficiency spectrum indicates a ten-shot-mean of 16.4~pC/shot and 0.83\% efficiency into the sub-beam. It is worth noting that the laser-specular electron spray is clipped by the OAP hole to create this sub-beam, and therefore the total accelerated $>$MeV charge/shot into the \emph{full} electron spray is almost certainly higher than these quoted numbers (consistent with the observations of \etal{Morrison}\citet{morrison2015backward}).

From an applications point-of-view, two notes on this work should be made. First, this setup creates a high peak current density of MeV electrons with a large \emph{average} current using a relatively inexpensive target setup in a rough vacuum and without stringent constraints on the laser pre-pulse contrast. This demonstration opens the door to a ultra-intense laser-based compact MeV gamma ray generator. In fact, using this setup, but without a hole in the OAP, one can easily generate highly directional rad/hr hard x-ray and $\gamma$-ray radiation from electron scattering with the OAP \cite{morrison2015backward}. Since the electrons are accelerated on fs-timescales and the fluence is high, this source could be used as a compact electron or X-ray backlighter to probe the temporal evolution of various systems of interest with a kHz cadence.
Second, the shot-to-shot fluctuations could be significantly reduced in future studies and in applications through modifying the experimental setup. Whereas in the current study the pre-plasma was created by the deliberate amplification and shaping of the natural laser ``noise'' of ASE and spurious short-pulse-replicas, one could instead introduce a separately amplified pulse in order to reduce shot-to-shot variation.
Another source of shot-to-shot variation is due to the convex curvature of the water column target. Because of this curvature, $\sim$1~\um{} transverse jitter in the laser focal spot position on the target changes the shot-to-shot effective depth of laser focus. This source of variation could be suppressed by changing the geometry of the flowing water from cylinder-like to sheet-like, and by widening the on-target pre-pulse diameter by defocusing any injected laser pre-pulse beam to increase transverse pre-plasma uniformity.

\section{Conclusion}
A direct measurement of kHz-pulsed super-ponderomotive electron acceleration in the back-reflection direction from normal incidence, 3~mJ laser interactions with solid-density liquid water at \Wcm[1.5]{18} has been performed and presented. The high-repetition-rate and low-laser-energy of this MeV~electron experiment is promising for electron-beam applications, and the results are scientifically novel in two ways: 1) the experiments were performed at normal incidence and the super-ponderomotive electrons were back-directed, details that required substantial modification of the experimental setup in order to study, and 2) the number of electrons/s.r./mJ (and electrons/s.r./minute) above 1.3~MeV exceeds all other studies for this laser energy and intensity. These measurements also directly confirm, for the first time, expectations from 2D and 3D PIC simulations that MeV electrons are indeed present in these interactions \cite{orban2015backward,ngirmang2016}. Since the experiment occurs at normal incidence and the ejection direction of super-ponderomotive electrons is laser specular, the mechanism is distinct from that of oblique-incidence electron acceleration. The number of electrons/MeV/s.r. detected at these energies also reinforces earlier conclusions that these ultra-intense laser interactions occur with a remarkably high conversion efficiency from laser energy to energetic electron energy \cite{morrison2015backward}. Methodologically, this work provides an unprecedented template for high-average-power HED experiment combining liquid-density targets, high acquisition rate plasma characterization, and large-sample statistical data analysis.
Future work will involve replacing the ASE/short-pulse pre-pulse with a controlled second-beam pre-pulse, co-timing acquisition from a variety of high-acquisition-rate laser/plasma diagnostics for statistical interrogations, and optimization of the electron beam divergence using adaptive laser optics.

\section*{Funding}

Air Force Office of Scientific Research (AFOSR). The  authors  acknowledge significant support from the Department of Defense High  Performance  Computing  Modernization  Program (DOD  HPCMP)  Internship  Program.    

\section*{Acknowledgments}
This research was sponsored by the Quantum and Non-Equilibrium Processes Division of the Air Force Office of Scientific Research, under the management of Dr. Enrique Parra, Program Manager. Support was also provided by the DOD HPCMP Internship Program. The Notre Dame Radiation Laboratory and J.A. L. are supported by the Division of Chemical Sciences, Geosciences and Biosciences, Basic Energy Sciences, Office of Science, United States Department of Energy through grant number DE-FC02-04ER15533. This contribution is NDRL-6000 from the Notre Dame Radiation Laboratory. ABW: 88ABW-2015-3607.



\newpage
\section*{References}
\bibliographystyle{unsrtnat}
\bibliography{bibfile}


\end{document}